\documentstyle[12pt] {article}
\input epsf.sty
\begin{document}
\begin{center}
{\Large\bf Interlayer Coupling in Magnetic/Pd Multilayers }\\

\bigskip
{\large Zhu-Pei Shi and Barry M. Klein}\\

{\em Department of Physics, University of California,
Davis, CA 95616}

\end{center}
\bigskip
\begin{center}
{\bf Abstract}\\
\begin{minipage}[t] {13.0cm}
The Anderson model of local-state conduction electron mixing is applied to the
problem of interlayer magnetic coupling in metallic
multilayered structures with palladium (Pd) spacer layers.
An oscillation period of $5$ spacer monolayers and the tendency towards
ferromagnetic bias of the interlayer magnetic coupling that we obtain
are consistent with the experimental data.
\end{minipage}
\end{center}

\vspace*{1.0cm}

The discovery of oscillatating interlayer magnetic couplings between
ferromagnetic layers separated by a nonmagnetic metallic
spacer~\cite{parkin:90} and of the related giant magnetoresistance
effect~\cite{france:88}, has
stimulated a lot of experimental and theoretical activity.
It has been shown that the periods
of the coupling are related to the topology of the Fermi
surface of the spacer layers.
This interpretation has been confirmed by model and first-principles
calculations, and is also supported by experiments~\cite{hein:adv,hath:chap}.
There are, however, other aspects of the coupling,
{\em e. g.\/}, the bias (ferro- or antiferro-magnetic) of the interlayer
magnetic coupling, which have not been fully explained.

For Fe(001) layers separated by Pd(001) spacers of thickness between 4 and 12
ML the interlayer magnetic coupling is observed to have a strong ferromagnetic
bias as seen in the experiments~\cite{celi:91}. Above a $13~ML$ thickness of
the Pd spacer the coupling begins
to be antiferromagnetic.
Metallic Pd is believed to be near the threshold of becoming ferromagnetic.
The non-relativistic calculations of Moruzzi and Marcus~\cite{moru:89} and of
Chen {\em et al.\/}~\cite{chen:89} predicted the onset of ferromagnetism in fcc
palladium with a 5\% expanded lattice.
In a recent publication the ferromagnetic bias of the coupling in magnetic
multilayer structures with a Pd spacer is explained in terms of the Pd as an
almost ferromagnetic media~\cite{hein:adv}.
Alternatively, in this paper, we interpret this ferromagnetic bias to be a
consequence of a competition between RKKY-like and superexchange couplings,
with RKKY coupling being dominant.

The RKKY-like coupling comes from intermediate states which correspond to
spin excitations of the Fermi sea. States corresponding to electron-hole pair
production in
the Fermi sea, with an attendant spin-flip, contribute to the RKKY coupling
as~\cite{shi:prb};
\begin{equation}
j_{RKKY}({\bf q})=
\sum _{n_{1},n_{2},k}\;[\;\;
\frac{\mid V_{n_{1}k}\mid ^{2}\,\mid V_{n_{2}k'}\mid ^{2}}{(\varepsilon
_{n_{2}k'}-\varepsilon _{+})^{2}}\;
\frac{
\theta (\varepsilon _{F}-\varepsilon _{n_{1}k}) \theta
(\varepsilon _{n_{2}k'}-\varepsilon _{F})}{\varepsilon _{n_{2}k'}
-\varepsilon _{n_{1}k}} \;+\;c.c.\,]\; ,
\end{equation}
where $\theta $ is a step function, $\varepsilon _{F}$ is the Fermi energy,
${\bf k'} ={\bf k} +{\bf q} +{\bf G}$, ${\bf G}$ is a vector of
the reciprocal lattice,
$\varepsilon _{+}$ is the energy of the
 local impurity state,
and $V_{nk}$ represents the strength
of the $s-d$ mixing interaction~\cite{ande:61}.

The superexchange coupling arises from charge excitations
in which electrons from local states are promoted above the Fermi sea (one
from each layer) providing
a second contribution to the coupling
in addition to the RKKY coupling~\cite{shi:prb}:
\begin{equation}
j_{S}({\bf q})
=-\sum_{n_{1},n_{2},k}\;[\;
\frac{\mid V_{n_{1}k}\mid ^{2}\,\mid V_{n_{2}k'}\mid ^{2}}{(\varepsilon
_{n_{2}k'}-\varepsilon _{+})^{2}}\;
\frac{\theta (\varepsilon _{n_{1}k}-\varepsilon _{F}) \theta
(\varepsilon _{n_{2}k'}-\varepsilon _{F})}{\varepsilon _{n_{1}k}
-\varepsilon _{+}}\;+\;c.c.\,]\; .
\end{equation}

The real space coupling between two sheets of spins can be obtained by Fourier
transforming Eqs. (1) and (2)~\cite{shi:prb}, with the coupling in the real
space given by,
\begin{equation}
J_{t}(z)=\frac{a}{2\pi }\;\int
_{0}^{\infty}\;dq_{z}\;j(q_{z})\;cos\,(q_{z}z)\;\;\;,\end{equation}
where $a$ is a lattice constant, and $z$ is in the direction perpendicular to
the
magnetic layers. The sign is chosen so that positive $J_{t}(z)$ signifies
ferromagnetic coupling.

Using the Slater-Koster parameters~\cite{papa:94}, one can easily diagonalize
small
matrices (9x9 for a typical transition metal) to obtain the energy bands and
density of states (DOS) for fcc Pd.
The electron wave
function $\mid n,{\bf k}>$ is a Bloch state belonging to band $n$ and wave
vector
${\bf k}$, and is expressed as linear combinations of localized orbitals:
\begin{equation}
\mid n,{\bf k}>=\frac{1}{\sqrt{N}}\;\sum _{\nu }\;e^{i{\bf k}\cdot {\bf
R}_{\nu}}\;\,\sum _{i}\; a_{ni}({\bf k})\;u_{i}({\bf r}-{\bf R}_{\nu})\;\;\; ,
\end{equation}
where $N$ is the number of cells in the material considered, ${\bf R}_{\nu }$
is a lattice vector, $u_{i}({\bf r}-{\bf R}_{\nu })$ is the $ith$ orbital basis
function, and $a_{ni}({\bf k})$ is a (real) normalized eigenvector component
determined by diagonalization of the single-particle Hamiltonian.
We use a plausible approximation
$V_{n_{1}k}V^{*}_{n_{2}k'}=
V^{2}\, M^{*}_{n_{1}k,n_{2}k'}\,({\bf q})$~\cite{shi:prb},where the matrix
element is defined as $M_{n_{1}k,n_{2}k'}=<n_{1}{\bf k}\mid e^{i{\bf q}\cdot
{\bf r}}
\mid n_{2}{\bf k'}>$.
The explicit expression for the matrix element is
\begin{equation}
M_{n_{1}k,n_{2}k'}=
\sum _{\nu }\;e^{i{\bf k}\cdot {\bf R}_{\nu}}\;\,\sum _{i,j}\; a_{ni}({\bf
k})\;
a_{n_{2}j}({\bf k'})\;\int \, d{\bf r}\;
u_{i}({\bf r})\,e^{i{\bf q}\cdot {\bf r}}\,u_{j}({\bf r}-{\bf R}_{\nu})\;\;\; .
\end{equation}
The essential conditions for the simplication of this matrix was already
discussed by Callaway {\em et al.\/}~\cite{call:84}.
$u_{i}({\bf r})$ are approximated as  Clementi wave functions for the
d states~\cite{call:84}, and $a_{ni}({\bf k})$ can be related to
the Slater-Koster parameters in Ref.~[10].

We consider one local level below $\varepsilon _{F}$ for simplicity and set
$\varepsilon _{+}=\varepsilon _{F}-E_{h}$, where
$E_{h}$ is the energy required to promote an electron from an occupied
local magnetic impurity level to the Fermi level.
Based on the band structure of bulk paramagnetic Pd, we have calculated the
couplings $j_{RKKY}(q_{z})$ and $j_{S}(q_{z})$ with $E_{h}=0.08~Ry$, as shown
in Fig.~1.

\begin{figure}
\hskip 0.3cm
\epsfysize=8.6cm
\vskip 0.5cm
\caption{
RKKY-like coupling, $j_{RKKY}(q_{z})$, and superexchange coupling,
$j_{S}(q_{z})$,
 for Pd(001) spacers in reciprocal space calculated with $E_{h}=0.08~Ry$. $V$
is expressed
 in rydbergs.
}
\label{f-jq}
\end{figure}

The couplings in real space are plottted in Fig.~2. The dashed line,
dotted line and solid line are for RKKY-like, superexchange,
and RKKY + Superexchange couplings, respectively.
 We see that the superexchange interaction gives a small contribution to the
coupling,
 and the total coupling has a strong ferromagnetic bias. This tendency for a
ferromagnetic bias resembles the experimental observation in Fe/Pd(001)
trilayered structures~\cite{celi:91}.
The $5~ML$ oscillation period in the calculated interlayer magnetic coupling
$J(z)$, as shown in Fig.~2, corresponds to the peak at $q_{z}\simeq
0.4\,\frac{2\pi }{a}$ in $j_{RKKY}(q_{z})$. It agrees with
the experimental period of $4-5~ML$ in Fe/Pd/Fe(001)
for trilayered structures~\cite{celi:91}.

\begin{figure}
\hskip 0.3cm
\epsfysize=8.6cm
\vskip 0.5cm
\caption{
Interlayer magnetic coupling in the real space. The dashed line and
dotted line are for RKKY-like and superexchange
 couplings, respectively.
The solid line is for the total coupling (RKKY+superexchange).
}
\label{f-jz}
\end{figure}

\begin{figure}
\hskip 0.3cm
\epsfysize=8.6cm
\vskip 0.5cm
\caption{
Density of states for Pd.
}
\label{f-dos}
\end{figure}

\begin{figure}
\hskip 0.3cm
\epsfysize=8.6cm
\vskip 0.5cm
\caption{
The total interlayer magnetic coupling for a densities of states with
a peak below, but near, the Fermi surface. Ferromagnetic bias appears
in the coupling.
}
\end{figure}

\begin{figure}
\hskip 0.3cm
\epsfysize=8.6cm
\vskip 0.5cm
\caption{
Interlayer magnetic coupling as a function of Pd spacer thickness.
The solid line and filled circles are our calculations and the experim
ental
data observed at $T=77~K$~\cite{hein:94}, respectively.
The theoretical (experimental) results are refered to the left (right)
 scales.
}
\label{f-bias}
\end{figure}

We explain the result of ferromagnetic bias for multilayers with
Pd spacers as being due to the structure of the Pd DOS and the
location of the Fermi level, as shown in Fig.~3. In particular,
the fact that the Fermi level falls {\em above\/} a peak in the
DOS followed by a relatively smooth and structureless DOS,
results in a relatively small superexchange contribution above
$\sim 4~ML$, and leads to a ferromagnetic bias driven by the
large RKKY coupling.

To confirm our explanation of the cause of the ferromagnetic
bias we use a free-electron gas model which enables us to
obtain analytic results for the couplings.
In a previous study~\cite{shi:euro}, we showed that in the
free-electron gas approximation,
RKKY + Superexchange coupling resembles
pure RKKY coupling, but without any magnetic bias.
To illustrate the effect of a peak in the density of states on top
of a free electron-like background, we use a ``toy model" calculation
by  adding a
Lorentzian shaped peak to the DOS of the free electron gas,
\begin{equation}
D(E)=\sqrt{E}
+\frac{\sqrt{E_{F}}}{(E
/\varepsilon_{F} -p)^{2}+h^{2}}\;\;\; ,
\end{equation}
where the position of the peak is at $E_{p} =p~ \varepsilon_{F}$, and $h$
adjusts height of the peak (small $h$ corresponds to a large peak).
For example, by fixing the position
($p =0.9$, below the Fermi level) and increasing the height of the peak ({\em
e.g.\/},
$h=0.3$),
ferromagnetic bias occurs in the coupling, as shown in Fig.~4.
We noted that in multilayer structures with a $Cr$ spacer,
RKKY + superexchange coupling gives an antiferromagnetic bias
due to the structure of the DOS,  with a peak above, but near to the Fermi
level $\varepsilon _{F}$~\cite{shi:euro}. This is also
confirmed in our ``toy model" calculations.

The coupling observed in the experiments contains a bilinear exchange
interaction $J_{1}$ and a biquadratic exchange interaction $J_{2}$. In
conventional notation, $J_{exp}=J_{1}-2J_{2}$~\cite{hein:adv}.
Here, RKKY-like and superexchange couplings are contained in the bilinear
coupling, $J_{1}$,
and a positive $J_{2}$ favors a perpendicular magnetic coupling. The
ratio of $J_{2}/J_{1}$ observed in the multilayer structures with $Cr$ spacers
is about $0.3$ -- $0.5$~\cite{hein:94}.
Magnetic multilayer structures with Pd spacers also have a relatively large
biquadratic exchange interaction $J_{2}$~\cite{hein:adv}.
With a proper choice of positive $J_{2}$ ($J_{2}/J_{1}\approx 0.5$ at $13~ML$),
the
bias can switch to antiferromagnetic for spacer thickness greater than $13~ML$,
as is observed experimentally. One can see that our calculated interlayer
magnetic coupling
with Pd spacers can be used to explain the experimental data~\cite{celi:91} as
shown in Fig.~5.

In summary, our model calculation has been able to reproduce the two salient
features of the interlayer magnetic coupling in $Fe/Pd(001)$ multilayer
structures:
large but rapidly decreasing ferromagnetic bias, and a $5~ML$ oscillation
period. The ferromagnetic bias arises from the competition between the
RKKY-like and superexchange couplings due to the special features of the
palladium DOS: relatively large peak below, but near to the Fermi level, and a
small DOS above the Fermi level.

This research was supported by the University Research Funds of the University
of California at Davis.
Zhu-Pei Shi would like to thank Peter M. Levy and John L. Fry
for very useful discussions.


\bigskip
\begin{thebibliography}{99}
\bibitem{parkin:90}
S. S. P. Parkin, N. More and K. P. More, Phys. Rev. Lett. {\bf 64},
2304 (1990).

\bibitem{france:88}
M. N. Baibich {\em et al.\/}, Phys. Rev. Lett. {\bf 61}, 2472 (1988).

\bibitem{hein:adv}
B. Heinrich and J. F. Cochran, Adv. Phys. {\bf 42}, 523 (1993); and references
therein.

\bibitem{hath:chap}
K. B. Hathaway, in {\em Ultrathin Magnetic Structures\/}, Vol. II, edited by B.
Heinrich and J. A. C. Bland (Springer \& Berlin, 1994), Chap. 2.
\bibitem{celi:91}
Z. Celinski, B. Heinrich and J. F. Cochran, J. Appl. Phys. {\bf 70}, 5870
(1991).
\bibitem{moru:89}
V. L. Moruzzi and P. M. Markus, Phys. Rev. B {\bf 39}, 471 (1989).
\bibitem{chen:89}
H. Chen, N. E. Brener and J. Callaway, Phys. Rev. {\bf 40}, 1443 (1989).
\bibitem{shi:prb}
Z. P. Shi, P. M. Levy and J. L. Fry,
Phys. Rev. Lett. {\bf 69}, 3678 (1992).
\bibitem{ande:61}
P. W. Anderson, Phys. Rev. {\bf 124}, 24 (1961).
\bibitem{papa:94}
D. A. Papaconstantopoulos, ``Handbook of the Band Structure of Elemental
Solids", pp159-160, Plenum Press, New York (1986).

\bibitem{call:84}
J. Callaway {\em et al.\/}, Phys. Rev. B {\bf 28}, 3818 (1983).

\bibitem{shi:euro}
Z. P. Shi, P. M. Levy and J. L. Fry, Europhys. Lett. {\bf 26},
473 (1994);
Phys. Rev. B {\bf 49}, 15159 (1994).

\bibitem{hein:94}
B. Heinrich {\em et al.\/}, Mat. Res. Soc. Sym. Proc., {\bf 313}, 119 (1993).

\end{thebibliography}
\end{document}